\documentclass[12pt]{iopart}
\usepackage{mathptmx}
\usepackage{graphicx}

\begin{document}

\title{Lie algebras of order $F$ and extensions of the Poincar\'e algebra}
\author{M. Rausch de Traubenberg}
\address{Laboratoire de Physique Th\'eorique, CNRS UMR  7085, 
Universit\'e Louis Pasteur, F-67084 Strasbourg cedex, France}
\begin{abstract}
$F-$Lie algebras are natural generalisations of Lie algebras ($F=1$)
and Lie superalgebras ($F=2$).
We give finite dimensional
examples of $F-$Lie algebras
 obtained by an inductive process from Lie algebras
and Lie superalgebras. Matrix realizations of the $F-$Lie algebras constructed
in this way from 
${osp}(2|m)$ are given.
 We obtain  a non-trivial extension of the Poincar\'e
algebra by an In\"on\"u-Wigner contraction of a certain $F-$Lie algebras with
$F>2$.
\end{abstract}

\section{Introduction}
Describing   the laws of physics in terms of  underlying symmetries has 
always been a powerful tool. Lie algebras and Lie superalgebras are 
central in particle physics, and the space-time symmetries can be obtained
by an In\"on\"u-Wigner contraction of certain Lie (super)algebras.
$F-$Lie algebras \cite{flie1,flie2,flie3}, 
a possible extension of Lie
(super)algebras,   have been considered some times ago as 
the natural structure underlying fractional supersymmetry  (FSUSY)
\cite{ flie1,durand,anyon,fsusyh} (one possible extension of supersymmetry).
In this contribution we show how one can construct many examples of
finite dimensional $F-$Lie algebras from Lie (super)algebras and
finite-dimensional 
FSUSY extensions of the Poincar\'e algebra are obtained by In\"on\"u-Wigner
contraction of certain $F-$Lie algebras.

\section{$F-$Lie algebras}
\label{Flie}

The natural mathematical  structure, generalizing the concept of
Lie superalgebras and relevant for the algebraic description of fractional
supersymmetry was introduced in~\cite{flie1}  and  called
an $F-$Lie algebra. We do not
want  to go into the  detailed definition of this structure here
and will  only recall the basic points, useful for our purpose.
More details can be found in~\cite{flie1}.

Let $F$ be a positive integer and $q=e^{2i \frac{\pi}{F}}$.
 We consider  now a complex vector space $S$ which has an automorphism
$\varepsilon$ satisfying $\varepsilon^F=1$.
We set ${  A}_k=S_{q^k}$, $1 \le k \le F-1$ and ${  B}=S_1$
($S_{q^k}$ is the eigenspace corresponding to the eigenvalue $q^k$ of
$\varepsilon$). Hence,
\[
S= B \oplus A_1 \oplus \cdots \oplus A_{F-1}.
\]
We say that $S$ is an $F-$Lie algebra if:
\begin{enumerate}
\itemsep0mm
\item $B$, the zero graded  part of $S$,   is a Lie algebra.
\item $A_i$ $(i=1,\ldots, F-1)$,  the $i$ graded part of $S$,
is a   representation of $B$.
\item  There are symmetric multilinear $B-$equivariant maps
\[
\left\{~, \ldots,~ \right\}:  \ {\cal S}^F\left(A_k\right)
\rightarrow B,
\]
 where  ${ \cal S}^F(D)$ denotes
the $F-$fold symmetric product
of $D$.
In other words, we  assume that some of the elements of the Lie algebra $B$ can
be expressed as $F-$th order symmetric products of
``more fundamental generators''.

\item The generators of $S$ are assumed to satisfy  Jacobi identities
($b_i \in B$, $a_i \in A_k$, $1 \le k \le F-1$):
\begin{eqnarray}
\left[\left[b_1,b_2\right],b_3\right] +
\left[\left[b_2,b_3\right],b_1\right] +
\left[\left[b_3,b_1\right],b_2\right] =0, \nonumber \\
\left[\left[b_1,b_2\right],a_3\right] +
\left[\left[b_2,a_3\right],b_1\right] +
\left[\left[a_3,b_1\right],b_2\right]  =0,\nonumber  \\
\left[b,\left\{a_1,\dots,a_F\right\}\right] =
\left\{\left[b,a_1 \right],\dots,a_F\right\}  +
\cdots +
\left\{a_1,\dots,\left[b,a_F\right] \right\}, \nonumber \\
\sum\limits_{i=1}^{F+1} \left[ a_i,\left\{a_1,\dots,
a_{i-1},
a_{i+1},\dots,a_{F+1}\right\} \right] =0. \label{rausch:eq:jac}
\end{eqnarray}
The  first three identities  are consequences of the previously
defined properties but the fourth is an extra constraint.

\end{enumerate}
More details  (unitarity,  representations, {\it etc.})
can be found in~\cite{flie1,flie3}. Let us first note that no relation between
different graded sectors is  postulated.
Secondly, the sub-space  $B \oplus A_k \subset S$ $(k=1, \ldots, F-1)$
is itself an
$F-$Lie algebra. From now on, $F-$Lie algebras of the types $B \oplus A_k $
will be considered.

Most of the examples of $F-$Lie algebras are infinite dimensional
(see {\it e.g.} \cite{flie1,anyon}). However in \cite{flie3} an inductive
theorem to construct finite-dimensional $F-$Lie algebras was proven:\\

\noindent
{\bf Theorem 1} {\it 
Let $\ {g_0}$ be a Lie algebra and $\ {g}_1$ a representation
of $\ {g_0}$ such that    

(i) $S_1=\ {g_0} \oplus \ {g_1}$ is an $F-$Lie algebra of
order $F_1 \ge 1$ \footnote{Strictly speaking this theorem is not
valid for $F_1=1$. In this case the notion of
graded $1-$Lie algebra has to be introduced \cite{flie3}.
 $S=\ {g}_0 \oplus \ {g}_1$, is a graded $1-$Lie algebra if (i)
$\ {g}_0$ a Lie algebra and $\ {g}_1$
is a representation of $\ {g}_0$ isomorphic to the
adjoint representation, (ii)
 there is a $\ {g}_0-$ equivariant map
$\mu :\ {g}_1 \to \ {g}_0$ such that
$\left[f_1, \mu(f_2) \right] + \left[f_2, \mu(f_1) \right] =0,
f_1,f_2 \in \ {g}_1$.
}; 

(ii) $\ {g_1}$ admits  a $\ {g_0}-$equivariant symmetric form
$\mu_2$ of order $F_2 \ge 1$.

\noindent
Then  $S=\ {g_0}  \oplus \ {g_1}$ admits 
an  $F-$Lie algebra structure of order $F_1+F_2$, which we call the
$F-$Lie algebra induced from $S_1$ and $\mu_2$.}\\

By hypothesis, there exist 
 $ {g_0}-$equivariant maps
 $\mu_1:{\cal S}^{F_1}\left( {g_1}\right) \longrightarrow 
 {g}_0$  and 
 $\mu_2:{\cal S}^{F_2}\left( {g_1}\right) \longrightarrow 
 \mathbf C$. 
Now, consider $\mu:{\cal S}^{F_1+ F_2}\left( {g_1}\right) 
\longrightarrow  {g}_0 \otimes \mathbf C \cong  {g}_0$ defined 
by 

\begin{eqnarray}
\label{eq:tensor}
&& \hskip 4truecm
\mu(f_1,\cdots,f_{F_1+F_2})=  \\
&&\frac{1}{F_1 !}\frac{1}{F_2 !} \sum \limits_{\sigma \in S_{F_1 + F_2}}
\mu_1(f_{\sigma(1)},\cdots,f_{\sigma({f_{F_1}})}) \otimes 
\mu_2(f_{\sigma(f_{F_1+1})},\cdots,f_{\sigma(f_{F_1+F_2})}),
\nonumber
\end{eqnarray}

\noindent
where $f_1,\cdots,f_{F_1+F_2} \in  {g_1}$ and 
$S_{F_1 + F_2}$ is the group of permutations on $F_1 + F_2$ elements.
By construction, this is a $ {g_0}-$equivariant map
from ${\cal S}^{F_1+ F_2}\left( {g_1}\right) \longrightarrow
 {g_0}$, 
thus the three first Jacobi identities are  satisfied.  The last Jacobi
identity,  is more difficult to check and  is a consequence of
the corresponding identity
 for the $F-$Lie algebra $S_1$ and  a factorisation property (see \cite{flie3}
for more details).

\section{Finite dimensional $F-$Lie algebras}
An interesting consequence of the theorem of the previous section is
that it enables us to construct an $F-$Lie algebras associated to
{\it any} Lie (super)algebras.

\subsection{Finite dimensional $F-$Lie algebras associated to Lie algebras} 

Consider the graded $1-$Lie algebra
$S= {g}_0 \oplus  {g}_1$
where $ {g}_0$ is a  
Lie algebra,  $ {g}_1$ is the adjoint representation of 
$ {g}_0$ and $\mu :  {g}_1 \to  {g}_0$ is the
identity. Let $J_1,\cdots, J_{\mathrm{dim} {g}_0}$ be
a basis of $ {g}_0$, and $ A_1,\cdots,
 A_{\mathrm{dim} {g}_0}$ the corresponding basis of $ {g}_1$.
The graded $1-$Lie algebra structure on $S$ is then:

\begin{eqnarray}
\label{eq:1-lie}
\left[J_a, J_b \right] = f_{ab}^{\ \ \ c} J_c, \qquad
 \left[J_a, A_b \right] = f_{ab}^{\ \ \ c} A_c, \qquad
\mu(A_a)= J_a,
\end{eqnarray}

\noindent
where $f_{ab}^{\ \ \ c} $ are the structure constants of $ {g}_0$,
The second ingredient to construct an $F-$Lie algebra is to 
define a symmetric invariant form on $g_1$.
But  on $g_1$, the adjoint representation of $g_0$, the invariant
symmetric forms are well known and correspond to the Casimir operators
\cite{ec}.
Then, considering 
a Casimir operator of order $m$ of $ {g}_1 \cong  {g}_0 $,  
 we can induce the
structure of an $F-$Lie algebra of order $m+1$ on 
$S_{m+1}=  {g}_0 \oplus  {g}_1$.
One can give explicit formulae for the bracket of these $F-$Lie
algebras as follows.
Let $h_{a_1 \cdots a_{m}}$ be a  Casimir operator
of order $m$ (for $m=2$, the Killing form
 $g_{ab}=\mathrm{Tr}(A_a A_b)$ is a primitive Casimir of
order two). 
Then, the $F-$bracket of the  $F-$Lie algebra is 

\begin{eqnarray}
\label{eq:mi-lie}
\left\{A_{a_1}, A_{a_2}, \cdots, A_{a_{m+1}} \right\} =
\sum \limits_{\ell =1}^{m+1}
h_{a_1 \cdots a_{\ell-1} a_{\ell +1} \cdots a_{m+1}} J_{a_\ell}
\end{eqnarray}

\noindent
For the Killing form  this gives

\begin{eqnarray}
\label{eq:3-lie}
\left\{A_a, A_b, A_c \right\} =
g_{ab} J_c + g_{ac} J_b + g_{bc} J_a.
\end{eqnarray}         

If ${g}_0= {sl}(2)$, the $F-$Lie algebra of
order three induced from the Killing form is the $F-$Lie algebra
of \cite{ayu}. 

\subsection{Finite dimensional $F-$Lie algebras associated to Lie 
superalgebras} 
The construction of $F-$Lie algebras associated to Lie superalgebras
is more involved. We just give here a simple example (for more details
see \cite{flie3}): the  $F-$Lie algebra of order $4$ $S=g_0 \oplus g_1$ 
 induced from the (i) Lie superalgebra 
$ {osp}(2|2m)= \left( {so}(2) \oplus  {sp}(2m)\right)
\oplus {  \mathbf C}^2 \otimes {  \mathbf C}^{2m},$
and (ii) the quadratic form $\varepsilon \otimes \Omega$, where 
$\varepsilon$ is the invariant symplectic form on   ${  \mathbf C}^2$
and $\Omega$ the   invariant symplectic form on   ${ \mathbf  C}^{2m}$.
Let $\left\{S_{\alpha \beta} = S_{\beta \alpha }
\right\}_{\begin{tiny}\begin{array}{l} 
1 \le \alpha \le 2m \\ 1 \le \beta \le 2 m
\end{array}\end{tiny}}$  be a basis of $ {sp}(2m)$ and 
$\left\{h \right\}$ be a basis of  $ {so}(2)$.
Let $\left\{F_{q \alpha}\right\}_{\begin{tiny}\begin{array}{l} 
q=\pm 1\\ 1 \le \alpha \le 2m
\end{array}\end{tiny}}$
be a basis of ${   \mathbf C}^2 \otimes { \mathbf   C}^{2m}$.
Then the four brackets of $S$ take the following form

\begin{eqnarray}
\label{eq:flie-orth}
\left\{F_{q_1  \alpha_1}, F_{q_2  \alpha_2}, F_{q_3 \alpha_3},
F_{q_4  \alpha_4} \right\}&=&  
\varepsilon_{q_1 q_3}  \Omega_{\alpha_1  \alpha_3}
\left(\delta_{q_2 + q_4}  S_{\alpha_2 \alpha_4} + 
\varepsilon_{q_2 q_4}   \Omega_{\alpha_2 \alpha_4} h
\right) \nonumber \\
 &+& \mathrm {perm.}  
\end{eqnarray} 

It is interesting to notice that this $F-$Lie algebra admits a simple
matrix representation \cite{flie3}:
$g_0=\left\{ \left(
\begin{array}{lll}
q&0&0 \cr
0&-q&0 \cr
0&0&S 
\end{array} \right), q\in {  \mathbf C}, S \in  {sp}(2n) \right\} 
\cong   {so}(2) \oplus   {sp}(2n)$
and $g_1= \left\{
\left(\begin{array}{lll}
0&0&F_+ \cr
0&0&F_- \cr
-\Omega F_-^t&-i \Omega F_+^t&0 
\end{array}\right), F_\pm \in {\cal M}_{1,2n}\left({  \mathbf C}\right)
 \right\}$.

\section{Finite-dimensional FSUSY extensions of the Poincar\'e algebra}

It is well known that supersymmetric extensions of the Poincar\'e
algebra can be obtained by  In\"on\"u-Wigner contraction
of certain Lie superalgebras. In fact, one can also obtain some 
 FSUSY extensions
of the Poincar\'e algebra  by In\"on\"u-Wigner contraction of 
certain $F-$Lie algebras  as we now show  with one
example \cite{flie3}. Let
$S_3= {sp}(4) \oplus \mathrm{ad} \   {sp}(4)$
be the real $F-$lie algebra of order three 
induced from the real graded $1-$Lie
algebra $S_1= {sp}(4) \oplus \mathrm{ad}  \  {sp}(4)$
and the Killing form on $\mathrm {ad} \ { sp}(4)$ (see eq. \ref{eq:3-lie}).
Using vector indices of ${so}(1,3)$ coming from the inclusion
${so}(1,3) \subset   {so}(2,3) 
\cong  {sp}(4)$, the bosonic part of
$S_3$ is generated by $M_{\mu \nu}, M_{\mu 4}$, with 
$\mu, \nu =0,1,2,3$ and the graded part by $J_{\mu \nu}, J_{4 \mu}$.
Letting $\lambda \to 0$
after  the In\"on\"u-Wigner contraction,

\begin{eqnarray}
\begin{array}{ll}
M_{\mu \nu} \to L_{\mu \nu},& M_{\mu 4} \to \frac{1}{\lambda} P_\mu 
\\
J_{\mu \nu} \to \frac{1}{\sqrt[3]{\lambda}} Q_{\mu \nu},& 
J_{4 \mu} \to \frac{1}{\sqrt[3]{\lambda}} Q_{\mu},
\end{array}
\end{eqnarray}

\noindent   
one sees that  
$L_{\mu \nu }$ and $P_\mu$ generate the $(1+3)D$ Poincar\'e
algebra and that $Q_{\mu \nu}, Q_\mu$ are the fractional supercharges
 in  respectively  the adjoint and vector representations of
$ {so}(1,3)$.
This $F-$Lie algebra of order three is therefore a non-trivial
extension of the Poincar\'e algebra where translations are cubes
of more fundamental generators. The subspace  generated by 
$L_{\mu \nu}, P_\mu, Q_\mu$ is also an $F-$Lie algebra of order three
extending the Poincar\'e algebra  in which
the trilinear symmetric brackets have the simple form: 

\begin{eqnarray}
\left\{Q_\mu, Q_\nu, Q_\rho \right \}=
\eta_{\mu \nu} P_\rho +  \eta_{\mu \rho} P_\nu + \eta_{\rho \nu} P_\mu,
\end{eqnarray}

\noindent
where $\eta_{\mu \nu}$ is  the Minkowski metric.

\section{Conclusion}
In this paper a sketch of the construction of $F-$Lie 
algebras associated to Lie (super)algebras were given. More complete
results, such as 
 a criteria for simplicity, 
representation theory, matrix realizations {\it etc.}, was  given 
in \cite{flie3}. 


\section*{References}
\begin{thebibliography}{9}
\normalsize
%
\bibitem{flie1}
Rausch de Traubenberg M and Slupinski M.~J. 2000  J. Math. Phys  41
 4556-4571 [hep-th/9904126].
%
%
\bibitem{flie2}
Rausch de Traubenberg M~ and Slupinski M.~J.~ 2002
{\it Proceedings of Institute of Mathematics of NAS of Ukraine}, 
p 548-554, Vol. 43,
Editors A.G. Nikitin, V.M. Boyko and R.O. Popovych, Kyiv, Institute of 
Mathematics [arXiv:hep-th/0110020].
\bibitem{flie3}
Rausch de Traubenberg  M~ and Slupinski  M.~J.~ 2002
{\it Finite-dimensional Lie algebras of order $F$},
arXiv:hep-th/0205113, to appear in   J. Math. Phys.
%
\bibitem{durand}
Durand S  1993 Mod. Phys. Lett  A 8 2323--2334 [hep-th/9305130].
%
\bibitem{anyon}
Rausch de Traubenberg M  and Slupinski M.~J.~ 1997
Mod. Phys. Lett.  A 12  3051-3066 [hep-th/9609203].
%
\bibitem{fsusyh}
Rausch de Traubenberg M 1998 hep-th/9802141 (Habilitation Thesis, in French).
%
\bibitem{ec}
Chevalley C~ and Eilenberg S  1948 Trans. Amer. Math. Soc.  63  85-124.
%
\bibitem{ayu}
Ahmedov H,  Yildiz A and   Ucan Y 2001  J.\ Phys.   A 34  6413-6424
[math.rt/0012058].
%

\end{thebibliography}
\end{document}